\documentclass[journal]{IEEEtran}
\usepackage{blindtext}
\usepackage{graphicx}
\usepackage[utf8]{inputenc}
\usepackage{graphics} % for pdf, bitmapped graphics files
\usepackage{epsfig} % for postscript graphics files
\usepackage{mathptmx} % assumes new font selection scheme installed
\usepackage{amsmath} % assumes amsmath package installed
\usepackage{amssymb}  % assumes amsmath package installed
\usepackage{cite}
\usepackage{multicol}
\usepackage{mathtools, cuted}
\usepackage[cmintegrals]{newtxmath}
\usepackage{epstopdf}
\usepackage{graphics} % for pdf, bitmapped graphics files
\usepackage{epsfig} 

\usepackage{mathptmx} % assumes new font selection scheme installed
\usepackage{cite}
\usepackage{multicol}
\usepackage{mathtools, cuted}
\usepackage[cmintegrals]{newtxmath}
\usepackage{mathtools}

\usepackage{mathptmx}
\usepackage{helvet}
\usepackage{courier}
\usepackage{subfigure}
\usepackage{type1cm}
\usepackage{titlesec}
\usepackage{epsfig} % for postscript graphics files
\usepackage{mathptmx} % assumes new font selection scheme installed
\usepackage{amsmath} % assumes amsmath package installed
\usepackage{amssymb}  % assumes amsmath package installed
\usepackage{cite}
\usepackage{multicol}
\usepackage{mathtools, cuted}
\usepackage[cmintegrals]{newtxmath}
\usepackage{makeidx}         % allows index generation
\usepackage{algorithmic}
\DeclareMathOperator*{\argmin}{\arg\!\min}

\usepackage{multicol}        % used for the two-column index
\usepackage[bottom]{footmisc}% places footnotes at page bottom
\usepackage{caption}
\usepackage{nomencl}
\usepackage[usenames, dvipsnames]{color}
\usepackage{algorithmic}
\usepackage{multicol}        % used for the two-column index
\usepackage[bottom]{footmisc}% places footnotes at page bottom
\usepackage{caption}
\usepackage{titlesec}
\usepackage{nomencl}
\usepackage[usenames, dvipsnames]{color}
\usepackage{multirow}

\usepackage{amsthm}
 
\theoremstyle{definition}
\newtheorem{definition}{Definition}

\newtheorem{remark}{Remark}

\newif\ifdraft
\draftfalse

\newcommand{\jb}[1]{{\ifdraft\color{cyan}[JB: #1]}\fi}

\begin{document}
\title{A Physics-Based Finite-State Abstraction for \\ Traffic Congestion Control}

\author{Hossein~Rastgoftar and Jean-Baptiste Jeannin
        % <-this % stops a space
\thanks{The authors are with the Department
of Aerospace Engineering, University of Michigan, Ann Arbor,
MI, 48109 USA e-mail: {hosseinr}@umich.edu.}%
% \thanks{A. Girard is with the Department
% of Aerospace Engineering, University of Michigan, Ann Arbor,
%  MI, 48109 USA e-mail: anouck@umich.edu.
% }%
% <-this % stops a space
}

% \and
% \IEEEauthorblockN{James Kirk\\ and Montgomery Scott}
\markboth{}%
{Shell \MakeLowercase{\textit{et al.}}: Bare Demo of IEEEtran.cls for Journals}



\maketitle

\begin{abstract}
% This paper presents traffic coordination problem by a finite-state abstraction an integrative data-driven physics-inspired approach to model and control traffic congestion in a resilient and efficient manner. While existing physics-based approaches commonly assign density and flow traffic states by using the Fundamental Diagram, this paper specifies the flow-density relation using past traffic information recorded in a time sliding window with a constant horizon length. 
This paper offers a finite-state abstraction of traffic coordination and congestion in a network of interconnected roads (NOIR). By applying mass conservation, we model traffic coordination as a Markov process. Model Predictive Control (MPC) is applied to control traffic congestion through the boundary of the traffic network. The optimal boundary inflow is assigned as the solution of a constrained quadratic programming problem. Additionally, the movement phases commanded by traffic signals are  determined using receding horizon optimization. In simulation, we show how traffic congestion can be successfully controlled through optimizing boundary inflow and movement phases at traffic network junctions. 
\end{abstract}
% \section{Introduction}

\section{Introduction}
Urban traffic congestion management is an active research area, and physics-based modeling of traffic coordination has been extensively studied by researchers over the past three decades. 
%  The existing physics-based approaches heavily rely on models.
 It is common to spatially discretize a network of interconnected roads (NOIR)
%  \jb{You need to define what a NOIR is and give some intuition}\hrs{NOIR was already defined in the abstract and intro}\jb{You only gave the meaning of the acronym. But what is a NOIR used for? Why is it useful? When was it introduced (which paper)? NOIR needs to be motivated and explained in a few sentences beyond just the meaning of the acronym.} 
 using the Cell Transmission Model (CTM) which applies mass conservation to model traffic coordination \cite{daganzo1995cell, gomes2006optimal}. To control and analyze traffic congestion, the Fundamental Diagram is commonly used to assign a flow-density relation at every traffic cell. While the Fundamental Diagram can successfully determine the traffic state for small-scale urban road networks, it may not properly function for congestion analysis and control in large traffic networks. Modeling of backward propagation, spill-back congestion, and shock-wave propagation is quite challenging. The objective of this paper is to deal with these traffic congestion modeling and control challenges.  In particular, this paper contributes a novel integrative data-driven physics-inspired approach to \textit{obtain a microscopic data-driven traffic coordination model} and \textit{resiliently control congestion in large-scale traffic networks}.

%  \subsection{Related Work}
Researchers have proposed light-based  and physics-based control approaches to address traffic coordination challenges. Fixed-cycle control is the traditional approach for the operation of traffic signals at intersections. The traffic network study tool
\cite{robertson1969transyt, tiwari2008continuity} is a standard fixed-cycle control tool for optimization of the signal timing.  Balaji and Srinivasan~\cite{balaji2011type} and Chiu~\cite{chiu1992adaptive} offer fuzzy-based signal control approaches to optimize the green time interval at junctions.  Physics-based traffic coordination approaches commonly use the Fundamental Diagram to determine traffic state (flow-density relation) \cite{zhang2012ordering, zhang2011transitions}, model dynamic traffic coordination \cite{han2012continuous}, incorporate spillback congestion \cite{gentile2007spillback, adamo1999modelling}, infuse backward propagation \cite{gentile2015using, long2008urban} effects into traffic simulation, or specify the feasibility conditions for traffic congestion control. 
Jafari and Savla~\cite{jafari2018structural} propose first order traffic dynamics inspired by mass flow conservation, dynamic traffic  assignment \cite{peeta2001foundations, janson1991dynamic}, and a cell transmission model \cite{daganzo1995cell, daganzo1994cell} to model and control freeway traffic coordination.  Model predictive control (MPC) is an increasingly popular approach for model-based traffic coordination optimization \cite{lin2012efficient, jamshidnejad2018sustainable, tettamanti2014robust}. Baskar et al.~\cite{baskar2012traffic} apply MPC to determine the optimal platooning speed for automated highway systems (AHS). Furthermore, researchers have applied fuzzy logic \cite{kammoun2014adapt, collotta2015novel, pau2018smart, yusupbekov2016adaptive}, neural networks \cite{moretti2015urban, tang2017improved, akhter2016neural, kumar2015short}, Markov Decision Process (MDP) \cite{ong2016markov, haijema2008mdp}, formal methods \cite{coogan2017formal, coogan2015traffic} and mixed nonlinear programming (MNLP) \cite{christofa2013person} for model-based traffic management.  Optimal control \cite{jafari2018structural, wang2018dynamic} approaches have also been proposed. {Rastgoftar et al.~\cite{DSCC2020} model traffic coordination as a probabilistic process where traffic coordination is controlled only through boundary inlet nodes. }
% \jb{Why do you not cite your DSCC paper? Seems very related, and it would be useful to point out explicitly the differences with this paper}
% In Ref. \cite{christofa2013person}, the responsive traffic applying mixed nonlinear programming (MNLP) is suggested to minimize the person delay. 
% Time-efficient and energy efficient traffic signal control methods was proposed in \cite{osorio2015energy}.
% Traffic signal control using model-free reinforcement learning \cite{mannion2016experimental, li2016traffic, abdoos2011traffic, arel2010reinforcement, abdulhai2003reinforcement, bingham2001reinforcement, wiering2000multi} has been offered to adaptively mange traffic coordination in a transportation infrastructure. 

% \subsection{Contribution}
This paper studies the problem of traffic coordination and congestion control in a network of interconnected roads (NOIR). We model traffic coordination as a mass conservation problem governed by the continuity partial differential equation (PDE). Through spatial and temporal discretization of traffic coordination, {this paper advances our previous work \cite{DSCC2020} by modeling} traffic {as} a Markov process controlled through ramp meters  (at boundary road elements) and traffic signals (at NOIR junctions). Given traffic feasibility conditions, MPC is applied to assign optimal boundary inflow such that traffic over-saturation is avoided at every NOIR road element. As the result, the optimal boundary inflow is continuously assigned as the solution of a constrained quadratic programming problem, and incorporated into traffic congestion planning. Given optimal boundary inflow, movement phase optimization is formulated as a receding horizon problem where discrete actions commanded by the traffic signals are assigned by minimization of  coordination costs over a finite time horizon. {Our proposed model ensures that traffic density is non-negative everywhere in the NOIR, if the traffic inflow is positive at every inlet boundary roads. Therefore, traffic coordination control can be commanded by a low computation cost.}

% \jb{How is this different (in a precise way) from all the related work you cite? Which papers are the closest? Why is it better?}\hrs{Happy to do that}\jb{Could you please answer this question since you're ``happy to do that''?}
% \section{Preliminaries}

% \jb{How do we validate this model? Why is this model correct?}

This paper is organized as follows. Preliminary notions of graph theory presented in Section \ref{Graph Theory Notions} are followed by traffic coordination modeling presented in Section \ref{Problem Statement}. Finite state abstraction of traffic coordination is presented in Section \ref{Problem Formulation}. Ramp-based and signal-based traffic congestion control is presented in Section \ref{Traffic Coordination Control}. Simulation results are presented in Section \ref{Simulation Results} followed by concluding remarks in Section \ref{Conclusion}.
\section{Graph Theory Notions}
 \label{Graph Theory Notions}
%   An NOIR is represented by a set of boundary nodes def by set $\mathcal{W}_B=\{1,\cdots,m_B\}$, set of junction nodes defined by set $\mathcal{W}_C=\{m_B+1,\cdots,m\}$, and set of roads defined by $\mathcal{R}=\left\{1,\cdots,m_R\right\}$.  Schematic of an NOIR consisting of $m_B=13$ boundary nodes, $17$
% junction nodes, and $m_R=53$ unidirectional roads is shown in Fig. \ref{NOIRExample}. As shown every unidirectional road is filled out by four road elements. 

\begin{figure*}[ht]
\centering
\subfigure[]{\includegraphics[width=.55\linewidth]{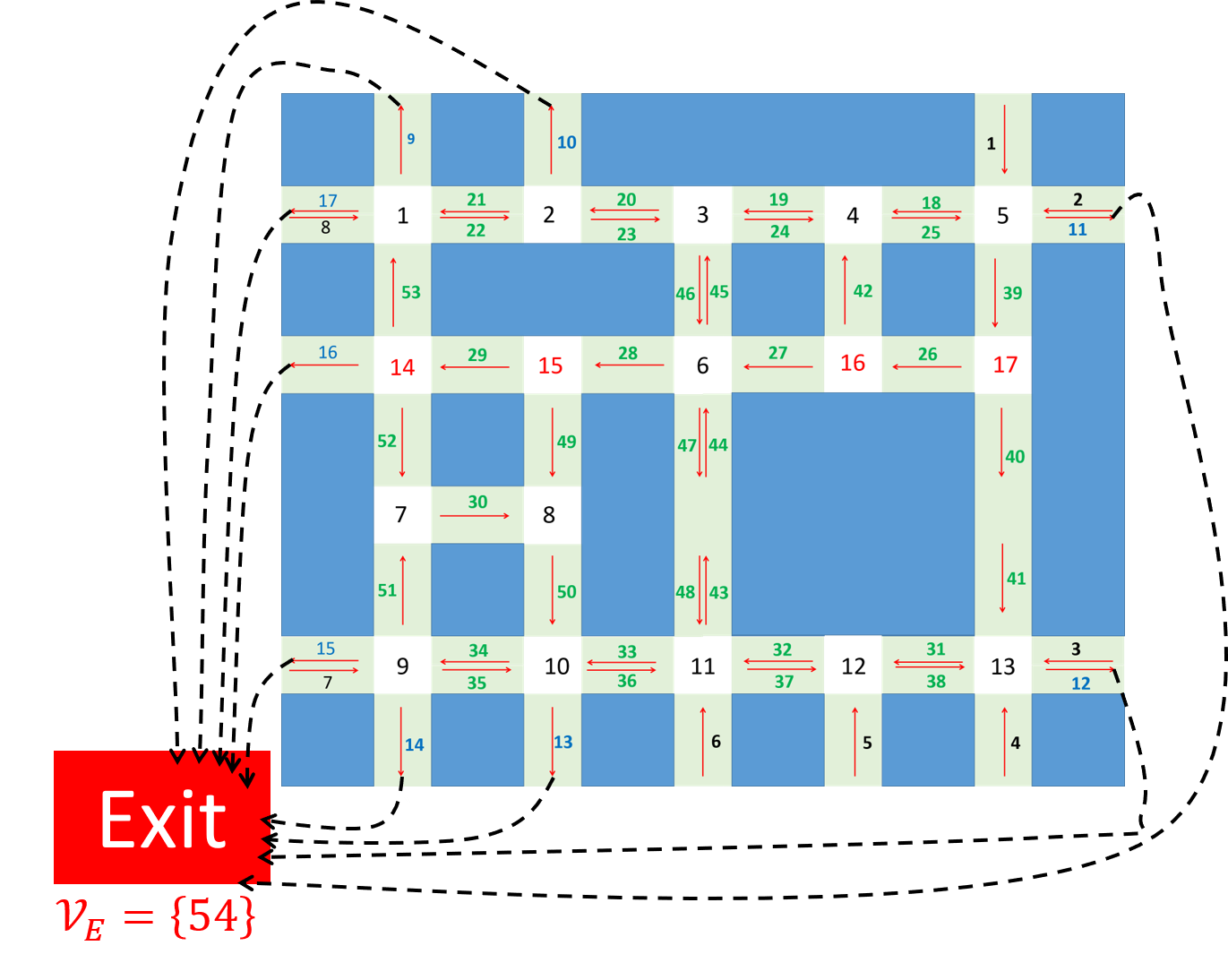}}
 \subfigure[]{\includegraphics[width=0.40\linewidth]{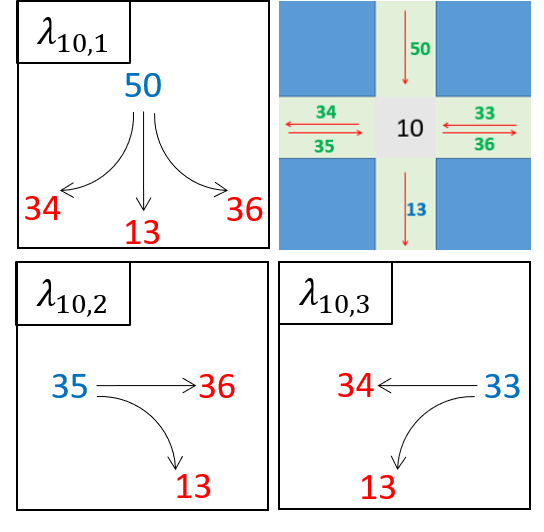}}
% \subfigure[]{\includegraphics[width=0.47\linewidth]{singlelane.jpg}}
%   \subfigure[]{\includegraphics[width=0.47\linewidth]{multilane.jpg}}
% \subfigure[$t=50$]{\includegraphics[width=0.45\linewidth]{falure50.jpg}}
% \vspace{-.4cm}
\caption{(a) Example NOIR with $53$ unidirectional roads. (b) Three possible movement phases at junction $10\in \mathcal{W}$. 
% \jb{This picture does not match your technical description. According to your technical description we have 3 incoming and 3 outgoing roads, hence we would have $\mu_{10}=3\times 3=9$, and\\
% $\lambda_{10,1}$ the edge from 50 to 34;\\
% $\lambda_{10,2}$ the edge from 50 to 13;\\
% $\lambda_{10,3}$ the edge from 50 to 36;\\
% $\lambda_{10,4}$ the edge from 35 to 36;\\
% $\lambda_{10,5}$ the edge from 35 to 13;\\
% $\lambda_{10,6}$ the edge from 33 to 34;\\
% $\lambda_{10,7}$ the edge from 33 to 13;\\
% $\lambda_{10,8}$ the edge from 33 to 36 (missing in your picture);\\
% $\lambda_{10,9}$ the edge from 35 to 34 (missing in your picture).\\
% Also your picture is missing the edges from 33 to 36 and from 35 to 34, which are included in your mathematical development (Equation (3)). If you wish to exclude them, you need to exclude them explicitly in the mathematical development. I don't see that.
% }
% \jb{If you disallow U-turns, this needs to be said explicitly in the paper}\hrs{Please see the last sentence right befor Section III.}
% % (c) $1$-D HDPI discretization of an NOIR road. (d) $2$-D HDPI doscretization of an NOIR road.
% % Block diagram of the LQG boundary controller design for a resilient macroscopic coordination.
% }
}
\label{NOIRExample}
\end{figure*}

Consider a NOIR with $m$ junctions defined by set $\mathcal{W}=\{1,\cdots,m\}$. An example of such a NOIR is shown in Fig.~\ref{NOIRExample}~(a). NOIR roads are identified by set $\mathcal{V}_R$ where $i\in \mathcal{V}_R$ is the index number of a road directed from an upstream junction to a downstream junction. Set $\mathcal{V}_R$ can be partitioned into a set of inlet boundary roads $\mathcal{V}_{in}$ and a set of non-inlet roads $\mathcal{V}_I$ such that
%\jb{TODO introduce $\mathcal{V}_{in}$ and $\mathcal{V}_I$ after the definitions below. Make V_E part of V_R. Don't make V_E a singleton from the beginning, just add that it is a singleton without loss of generality by connecting all exit nodes to V_E, and refer to the example.}
\begin{equation}
    \mathcal{V}_{R}=\mathcal{V}_{in}\bigcup \mathcal{V}_I.
\end{equation}
{
We also define a single ``Exit'' road defined by singleton $\mathcal{V}_E$. {Note that the ``Exit'' road does not represent a real road element (See Fig. \ref{NOIRExample} (a)); it is defined to model traffic coordination by a finite-state Markov process.}
We spatially discretize the NOIR using graph $\mathcal{G}\left(\mathcal{V},\mathcal{E}\right)$ with node set $\mathcal{V}=\mathcal{V}_R\bigcup \mathcal{V}_E$ and edge set $\mathcal{E}\subset \mathcal{V}\times \mathcal{V}$. Note that the nodes of graph $\mathcal{G}$ are the roads of our NOIR, and subsequently we use ``road'' and ``node'' interchangeably.
%\jb{TODO: clean up the use of ``road'' and ``node''; just use one of them}
Graph $\mathcal{G}$ is directed and the edge set $\mathcal{E}$ hold the following properties:
\begin{enumerate}
    \item{Traffic flow is directed from road $i$, if $(i,j)\in \mathcal{E}$. }
    \item{Real roads defined by set $\mathcal{V}_R$ are all unidirectional. Therefore, $(j,i)\notin \mathcal{E}$, if $(i,j)\in \mathcal{E}$.} 
    % \jb{Aren't roads defined by $\mathcal{V}_{R}$? I'm confused}\jb{Do you really mean that, or do you mean that $(i,j)\in \mathcal{E}$ does not necessarily imply $(j,i)\in \mathcal{E}$ (which is the definition of a directed graph)?}\jb{In figure 1, the roads of $\mathcal{V}_{R}$ are represented as arrows which means they are directed; but this is not present in the formalism, where only the edges of $\mathcal{E}$ are directed. I find that confusing. For example, could I have an edge from 50 to 34 (as you have) but also from 33 to 50 and from 36 to 50? Right now your formalism does not disallow that.}
\end{enumerate}

 Given graph $\mathcal{G}\left(\mathcal{V},\mathcal{E}\right)$, global in-neighbor, global out-neighbor, inlet boundary nodes, non-inlet nodes, and ``Exit'' node are formally defined as follows: % Given edge set $\mathcal{E}$, {the global in-neighbors and out-neighbors of road } $i\in \mathcal{V}$ are formally defined in Definitions \ref{Def1} and \ref{Def2} below.
\begin{definition}\label{Def1}
Given edge set $\mathcal{E}$, the global in-neighbors of road $i$ are defined by set
\begin{equation}
%i\in \mathcal{V},\qquad     
\mathcal{I}_{i}=\left\{j\in \mathcal{V}_R: (j,i)\in \mathcal{E}\right\}.
\end{equation}
\end{definition}
\begin{definition}\label{Def2}
Given edge set $\mathcal{E}$, the global out-neighbors of road $i$ are defined by set
\begin{equation}
%i\in \mathcal{V},\qquad     
\mathcal{O}_{i}=\left\{j\in \mathcal{V}: (i,j)\in \mathcal{E}\right\}.
\end{equation}
\end{definition}
% \begin{subequations}
% \begin{equation}
% %i\in \mathcal{V},\qquad     
% \mathcal{I}_{i}=\left\{j\in \mathcal{V}_R: (j,i)\in \mathcal{E}\right\},
% \end{equation}
% \begin{equation}
% %i\in \mathcal{V},\qquad     
% \mathcal{O}_{i}=\left\{j\in \mathcal{V}: (i,j)\in \mathcal{E}\right\}.
% \end{equation}
% \end{subequations}
\begin{definition}
Inlet boundary roads have no in-neighbors at any time, and they are formally defined by set 
\begin{equation}
    \mathcal{V}_{in}=\{i\in \mathcal{V}_R:\mathcal{I}_i=\emptyset \wedge \mathcal{O}_i\neq\emptyset\}.
\end{equation}
\end{definition}
%\jb{TODO Easier setup: just one set of roads, and for all roads either I_i or O_i is non-empty (otherwise it's isolated). If I_i is empty it's an inlet boundary node. If O_i is empty is't an exit node (we assume wlog that it's a singleton). If both I_i and O_i are non-empty it's a non-inlet road.}
\begin{definition}
Non-inlet roads have {at least one in-neighbor and one out-neighbor} at any time, 
% \jb{possible typo: can you please double-check this, I don't think you mean no out-neighbors for non-inlet roads} 
and they are formally defined by set 
\begin{equation}
    \mathcal{V}_{I}=\mathcal{V}_R\setminus \mathcal{V}_{in}.
\end{equation}
\end{definition}
\begin{definition}
The ``Exit'' node is formally defined as follows:
\begin{equation}
    \mathcal{V}_{E}=\{i\in \mathcal{V}:\mathcal{I}_i\neq\emptyset \wedge \mathcal{O}_i=\emptyset \}
\end{equation}
where we assume that $\mathcal{V}_{E}$ is a singleton.
% \jb{From this definition, how do you know that $\mathcal{V}_{E}$ is a singleton?}
\end{definition}
Without loss of generality, inlet boundary nodes are indexed from $1$ through $N_{in}$,  non-inlet roads are indexed from $N_{in}+1$ through $N$, and the ``Exit'' node is indexed by $N+1$.} Therefore 
$\mathcal{V}_{in}=\{1,\cdots,N_{in}\}$, $\mathcal{V}_I=\{N_{in}+1,\cdots,N\}$, and $\mathcal{V}_E=\{N+1\}$ define the inlet, non-inlet, and ``Exit'' nodes, respectively. 
% \jb{Note that at this point the reader does not understand what that is about.}
We use graph $\mathcal{G}\left(\mathcal{V},\mathcal{E}\right)$ to define interconnections between the NOIR roads, 
$
\mathcal{V}=\mathcal{V}_{R}\bigcup\mathcal{V}_E 
$
and $\mathcal{E}\subset \mathcal{V}\times \mathcal{V}$ define nodes and edges of graph $\mathcal{G}$. 
% define ``global in-neighbor set'' and ``global out-neighbor set'' of road $i\in \mathcal{V}$, respectively. Also, sets $\mathcal{V}_{in}$, $\mathcal{V}_I$, and $\mathcal{V}_E$ are formally defined as follows: 
%\jb{are those lemmas (requires proof) or definitions? If you're defining them in two different ways (you already defined them at the beginning of the section), you need to show that the two definitions match. I think it's reasonable to take this as definitions, but the way it's written is unclear. Also, if you take those as definitions, how do you ensure that $\mathcal{V}_E$ is a singleton?}
%\hrs{I formally defined them in Definitions 1 through 5. I did not understand your question. We do not need to provide proofs for definitions.}\jb{My point is, you've already defined $\mathcal{V}_{in}$, $\mathcal{V}_I$, and $\mathcal{V}_E$ at the beginning of the Section (just under Equation 1). Do the definitions match and why?}
% \begin{subequations}
% \begin{equation}
%     \mathcal{V}_{in}=\{i\in \mathcal{V}_R:\mathcal{I}_i=\emptyset \wedge \mathcal{O}_i\neq\emptyset\},
% \end{equation}
% \begin{equation}
%     \mathcal{V}_{I}=\mathcal{V}_R\setminus \mathcal{V}_{in},
% \end{equation}
% \begin{equation}
%     \mathcal{V}_{E}=\{i\in \mathcal{V}:\mathcal{I}_i\neq\emptyset \wedge \mathcal{O}_i=\emptyset\}.
% \end{equation}
% \end{subequations}

{The NOIR shown in Fig.~\ref{NOIRExample} contains $53$ unidirectional ``real'' roads identified by set $\mathcal{V}_R=\{1,\cdots,53\}$ and a virtual ``Exit'' node identified by set $\mathcal{V}_E=\{54\}$, i.e. $\mathcal{V}=\mathcal{V}_R\bigcup \mathcal{V}_E$.
%Because the ``Exit'' node does not represent a real road, it is not shown in Fig. \ref{NOIRExample}.
Note that roads $9,\cdots,17\in \mathcal{V}_I\subset \mathcal{V}_R$ are in-neighbors to the ``Exit'' node $54\in \mathcal{V}_E$, as represented by the dotted lines. Thus
\[
\mathcal{I}_{54}=\{9,\cdots,17\}.
\]
% \jb{I don't see 54 on the picture. Is it because it is virtual?}
Inlet nodes are identified by $\mathcal{V}_{in}=\{1,\cdots,8\}$ and  $\mathcal{V}_I=\{9,\cdots,53\}$ defines all non-inlet roads.
% \jb{So if I understand correctly road $13\in \mathcal{V}_I$. But given edges 35 to 13 and 33 to 13, it seems that $\mathcal{I}_{13}\neq \emptyset$, and I also believe $\mathcal{O}_{13} = \emptyset$ (I don't see what could be in there). Therefore according to your definition/lemma (3c), we also have $13\in \mathcal{V}_E$. Am I understanding correctly or am I missing something? The same should be true for roads 9, 10, 11, 12, 14, 15, 16 and 17.}\hrs{You are completely correct. I added some texts.}\jb{OK but then $\mathcal{V}_E$ is not a singleton anymore: it contains not only 54, but also 9, 10, 11, 12, 13, 14, 15, 16 and 17. Isn't that inconsistent with what you wrote before? Can you please double-check? Also, where exactly is the text you added?}
}

% {By defining $\mathcal{V}$ as $\mathcal{V}=\mathcal{V}_R\bigcup \mathcal{V}_E$, a finite-state abstraction of }

% \jb{Why do you choose $\mathcal{V}_E$ to be a singleton (which wasn't true in the DSCC paper)?}\hrs{$\mathcal{V}_E$ is not a real node. It is an imaginary added added to make }
% \jb{What do you mean by ``not a real road element''? Also see my long comment at the bottom of Figure 1.}

% \jb{Unless I'm misunderstanding something, in your Figure 1 $\mathcal{V}_E$ is not a singleton, but rather 
% $\mathcal{V}_E=\{9,10,11,12,13,14,15,16,17\}$. We need to have the description and the Figure match.}

% \jb{Define explicitly $\mathcal{V}$, $\mathcal{V}_{in}$, $\mathcal{V}_I$, $\mathcal{V}_E$ and $\mathcal{E}$ explicitly on the example of Figure 1, for the lost reader, just like we did on DSCC.}

% \jb{Roads are the vertices in your graph, is that right? I already made this comment for DSCC paper, but wouldn't it be better and much more intuitive to have junctions be vertices and roads be edges? Is there anything specific and technical that prevents us from doing that?}\hrs{No-Junctions are defined by set $\mathcal{W}$ as stated in the beginning of Section II. Junctions are not defined by set $\mathcal{V}$.}
% \jb{Right, that's not what you're doing. But wouldn't it be simpler and more intuitive?}

\textbf{Movement Phase Rotation:} 
{At each intersection, we define \emph{movement phases} representing the different possible configurations of traffic light states at that intersection or, equivalently, the different possible paths that are allowed at that intersection. For instance, in the example of Fig.~\ref{NOIRExample}, intersection number 10 has three lights -- at the ends of roads 33, 35 and 50 -- and three different movement phases:
\begin{itemize}
    \item the first movement phase $\lambda_{10,1}$ corresponds to a green light at the end of road 50, and red lights at the ends of roads 33 and 35; equivalently, cars are allowed to circulate from road 50 to roads 34, 13 or 36, and no other circulation is allowed;
    \item the second movement phase $\lambda_{10,2}$ corresponds to a green light at the end of road 35, and red lights at the end of roads 33 and 50; cars are only allowed to circulate from road 35 to either road 13 or 36;
    \item the third movement phase $\lambda_{10,3}$ corresponds to a green light at the end of road 33, and red lights at the end of roads 35 and 50 to be red; cars are only allowed to circulate from road 33 to either road 13 or 34.
\end{itemize}
Those three movement phases define the three possible configurations of the lights at intersection number 10, and over time the lights of intersection 10 alternatively go over those movement phases.}

{Formally, let $\mathcal{M}_{in,j}\subset \mathcal{V}_R$ define incoming roads and $\mathcal{M}_{out,j}\subset \mathcal{V}_R$ define outcoming roads at junction $j\in \mathcal{W}$. Every junction $j$ is associated with $\mu_j$ movement phases that can be commended by the traffic signals. The set of edges $\lambda_{j,k}\subset \mathcal{M}_{in,j}\times \mathcal{M}_{out,j} \subset \mathcal{E}$ is the $k$-th movement phase commanded at junction $j\in \mathcal{W}$  where $k=1,\cdots,\mu_j$.}
Movement phases at junction $j\in \mathcal{W}$ are defined by finite set $\mathbf{\Lambda}_j$ as follows: 
\begin{equation}
\label{movementphasedefinition}
  \mathbf{\Lambda}_j=\bigcup_{k=1}^{\mu_j}\{\lambda_{j,k}\}=\{\lambda_{j,1},...,\lambda_{j,\mu_j}\}
\end{equation}
where $j\in \mathcal{W}$ and $k=1,\cdots,\mu_j$. 
{Note that $\mathbf{\Lambda}_j$ is a set of {subsets of edge set $\mathcal{E}$}, i.e., is contained in the powerset of $\mathcal{E}$.}
% \jb{can you double-check this definition (7)? It seems weird to me but I'm not sure. Are you sure you do not mean $$\mathbf{\Lambda}_j=\bigcup_{k=1}^{\mu_j}\{\lambda_{j,k}\}=\{\lambda_{j,1},...,\lambda_{j,\mu_j}\}$$
% which is a set of sets of edges (rather than a set of edges), and is contained in the powerset of \mathcal{E} rather than contained in \mathcal{E} itself.}
%
% where $\mu_j$ is the total number of movement phases at junction $j\in \mathcal{W}$ and $\lambda_{j,i}\subset \mathcal{M}_{in,j}\times \mathcal{M}_{out,j} \subset \mathcal{E}$ define the $i$-th movement phase at junction $j\in \mathcal{W}$.
%\jb{Do you mean $\lambda_{j,i}\in \mathcal{E}$?}
 We can define 

% \jb{Are all those notations standard? Is there any way we can simplify them? Right now I find it difficult to follow all the notations. Maybe some intuition would help.}\hrs{I think they are all correct and concise. This is a novel formulation of the traffic congestion modeling and control so the formulation is novel. Perhaps you can rewrite formulations better.}
\begin{equation}
    \mathbf{\Lambda}=\mathbf{\Lambda}_1\times \cdots \times \mathbf{\Lambda}_m
\end{equation}
% \jb{The reader has no idea what you're trying to do since there is no intuition. Please provide some intuition.}
% \jb{Do we have $\mathbf{\Lambda}=\mathcal{E}$? If not was is also in $\mathcal{E}$ but not in $\Lambda$?}\hrs{No. $\mathcal{E}$ is a finite set of two-tuple members while $\mathbf{\Lambda}$ is a finite set of $m$-tuple members.}
 as the set of all possible movement phases across the NOIR.
% \begin{equation}
%      \mathbf{\Lambda}_j=\mathcal{M}_{in,j}\times \mathcal{M}_{out,j}\subset \mathcal{E}.
% \end{equation}
% define all possible movement phases 
% \jb{intuitively, what is a movement phase? I don't get the intuition. Also, where does this concept come from?}
% \jb{``A movement phase is a set of motions/edges activated by a traffic signal when a green light is active.'' You need to say it.}\hrs{I think (7) and (8) clearly define the movement phases. I also added some texts to make it more clear.}\jb{I can read the math but it gives me no intuition. What is a movement phase intuitively? I still have no idea.}
%at junction $j\in  \mathcal{W}$.
Transitions of movement phases are cyclic 
% \jb{I don't understand what this sentence means, what you are trying to do here or where you are going. Please provide some intuition on what is going on.} 
at every junction $j\in \mathcal{W}$, and defined by {cycle} graph $\mathcal{C}_j\left(\mathbf{\Lambda}_j,\mathbf{\Xi}_j\right)$ with node set $\mathbf{\Lambda}_j$ and edge set 
\begin{equation}
    \mathbf{\Xi}_j=\left\{\left(\lambda_{j,1},\lambda_{j,2}\right),\cdots,\left(\lambda_{j,\mu_j-1},\lambda_{j,\mu_j}\right),\left(\lambda_{j,\mu_j},\lambda_{j,1}\right)\right\}
\end{equation}
% \jb{Typo: shouldn't the last element of the set be $\left(\lambda_{j,\mu_j},\lambda_{j,1}\right)$?}
{Intuitively, first $\lambda_{j,1}$ is the active movement phase defining the current traffic light states and equivalent authorized paths at junction $j\in\mathcal{W}$; then the active movement phase is switched to $\lambda_{j,2}$, then to $\lambda_{j,3}$,..., then to $\lambda_{j,\mu_j}$, then back to $\lambda_{j,1}$ to restart the cycle.}

%\jb{TODO define the junctions with letters or roman numerals or something to avoid confusion}
{Fig.~\ref{NOIRExample}~(b) shows all possible movement phases at junction $10\in \mathcal{W}$ of the NOIR shown in Fig. \ref{NOIRExample} (a), where $\mathcal{W}=\{1,\cdots,13\}$ defines the junctions. The incoming and outcoming roads are defined by set $\mathcal{M}_{in,10}=\{33,35,50\}$ and $\mathcal{M}_{out,10}=\{13,34,36\}$, respectively. There are three movement phases $\lambda_{10,1}=\{(50,34), (50,13),(50,36)\}\subset \mathcal{E}$, $\lambda_{10,2}=\{(35,13), (35,36)\}\subset \mathcal{E}$, and $\lambda_{10,1}=\{(33,13), (33,34)\}\subset \mathcal{E}$. Note that U-turns are disallowed at every junction of the Example NOIR shown in Fig. \ref{NOIRExample}.}

\textbf{Movement Phase Activation Time:} It is assumed that movement phase $\lambda_{j,k}\in \mathbf{\Lambda}_j$ ($k=1,\cdots,\mu_j$)
% \jb{That does not make sense given your definition (7)}
cannot be active more that $T_{L,j}$ time steps, where $T_{L,j}\in \mathbb{N}$ is equivalent to $T_{L,{j}}\Delta T$ seconds, and $\Delta T$ is a known constant time step interval. 
% \jb{What is the intuition for providing such a restriction/bound?} 
{Because movement rotation is cyclic at every junction $j\in \mathcal{W}$, we define} the \textit{maximum activation time} $T_{L,j}$  for {every} movement phase at NOIR junction $j\in \mathcal{W}$. Define $T_{j}$ as the activation time of a movement phase at junction $j\in \mathcal{W}$, where $T_{j}\leq T_{L,j}$. Note that $T_{j}$ is independent of index $k\in \{1,\cdots,\mu_j\}$ and is counted from the start time of  a movement phase $\lambda_{j,k}$ at junction $j\in \mathcal{W}$. Given $T_{j}$ and $T_{L,j}$, we define \textit{activation index}
 \[
j\in \mathcal{W},\qquad
% ~\lambda_{j,i}\in \mathbf{\Lambda}_j,\qquad 
\tau_j=\displaystyle\left\lfloor\dfrac{T_{j}}{T_{L,j}}\right\rfloor\in \{0,1\}
 \]
%  \jb{what do those square brackets mean? Why is $\tau_j\in\{0,1\}?$ Is it the floor function? Isn't standard notation $\displaystyle\left\lfloor\dfrac{T_{j}}{T_{L,j}}\right\rfloor$?}
 at every intersection $j\in \mathcal{W}$, where $\lfloor\cdot\rfloor$ denotes the floor function. Because $T_{j}\leq T_{L,j}$, $\tau_j\in \{0,1\}$ is a binary variable  assigning whether the active movement phase must be overridden or not. If $\tau_j=0$, the current movement $\lambda_{j,k}$ ($k=1,\cdots,\mu_j,~j\in \mathcal{W}$) can still remain active. Otherwise, the active movement phase is overridden and the next movement phase must be selected.

%\jb{Where does this whole framework and modeling come from? Did you make it up? If yes, why do you not use an existing framework and standard notations; and which papers is it inspired by? If not, what are the sources?}\hrs{This is something novel and creative. If it is similar to the existing work then it is not novel.}

% \jb{How is this work different from the existing formulations, and why is it better from the existing frameworks? What other frameworks trying to do similar things exist?}\hrs{In the existing literature, signalized control is modeled as a utility optimization problem \cite{mehr2017joint}. This ppaper provides a novel approach for the joint signalized and boundary control of traffic congestion.}

The network movement phase is denoted by $\lambda=\left(\lambda_1,\cdots,\lambda_m\right)\in \mathbf{\mathbf{\Lambda}}$ where $\lambda_j\in \mathbf{\Lambda}_j$ and $j\in \mathcal{W}$. 
%\jb{TODO the switch of notations from $\lambda_{j,k}$ to $\lambda_j$ may be confusing}
We define the switching communication graph $\mathcal{G}_{\lambda}\left(\mathcal{V},\mathcal{E}_\lambda\right)$ to specify the inter-road connection under movement phase $\lambda\in \mathbf{\Lambda}$, where $\mathcal{E}_\lambda\subset \mathcal{E}$ defines the edges of graph $\mathcal{G}$.
{Per movement phase definition given in  \eqref{movementphasedefinition}, $\mathcal{E}_\lambda=\cup_{k=1}^m \lambda_k$.}
In-neighbors and out-neighbors of road (or Exit node) $i\in \mathcal{V}$ is defined by the following sets:
\begin{subequations}
\begin{equation}
i\in \mathcal{V},\lambda\in \mathbf{\Lambda},\qquad     \mathcal{I}_{i,\lambda}=\left\{j\in \mathcal{V}_R: (j,i)\in \mathcal{E}_\lambda\right\},
\end{equation}
\begin{equation}
i\in \mathcal{V},\lambda\in \mathbf{\Lambda},\qquad     \mathcal{O}_{i,\lambda}=\left\{j\in \mathcal{V}: (i,j)\in \mathcal{E}_\lambda\right\}.
\end{equation}
\end{subequations}
Given the above definitions, {for any $\lambda\in\mathbf{\Lambda}$, $\mathcal{I}_{i,\lambda}\subset \mathcal{I}_i$
and $\mathcal{O}_{i,\lambda}\subset \mathcal{O}_i$, thus:}
\begin{enumerate}
    \item{for every $\lambda\in \mathbf{\Lambda}$, in-neighbor set $\mathcal{I}_{i,\lambda}=\emptyset$ if $i\in \mathcal{V}_{in}$;}
    \item{for every $\lambda\in \mathbf{\Lambda}$, out-neighbor set $\mathcal{O}_{i,\lambda}=\emptyset$ if $i\in \mathcal{V}_E$.}
\end{enumerate}

% \jb{We need to refer to Figure 1 somewhere in the text. Right now you are not referring to Figure 1 anywhere.}

% \jb{What exactly is new in all this modeling you are doing? It is unclear from reading this Section.}\hrs{Please see the paragraph below.}

\section{Traffic Coordination Model}\label{Problem Statement}
We use the mass conservation law to model traffic at every NOIR road element $i\in \mathcal{V}$. Let $\rho_i$, $y_i$, and $z_i$ denote traffic density, traffic inflow, and traffic outflow at every road element $i\in \mathcal{V}$. 
%  \textit{\underline{Time discretization:}} It is assumed that NOIR movement phases can potentially change every $\Delta T$ seconds. Therefore, NOIR movement phase does not over the time interval $t\in [T_k,T_{k+1}]$, where $T_k=k\Delta T$ and $T_k=(k+1)\Delta T$. By discretizing time interval $[T_k,T_{k+1}]$, sample time $t_{(k-1)N_\tau+l}=T_k+{(l-1)\over N_\tau}\Delta T$ where $j=1,\cdots,N_\tau$. 
 Traffic dynamics governed by mass conservation is:
 \begin{equation}
     \rho_{i}\left(k+1\right)=\rho_{i}\left(k\right)+y_i\left(k\right)-z_i\left(k\right),
 \end{equation}
 where 
 \begin{subequations}
 \begin{equation}
     z_{i}\left(k\right)= 
     \begin{cases}
     p_{i}\left(\lambda\right)\rho_{i}\left(k\right)&i\in \mathcal{V}_R,~ \forall\lambda\in \mathbf{\Lambda}\\
     \rho_i\left(k\right)+y_i\left(k\right)&i\in \mathcal{V}_E,~ \forall \lambda\in \mathbf{\Lambda} \\
      \end{cases}
    %  \begin{cases}
    % %  v_{e_i}\left(k\right)&e_i\in \mathcal{V}_{A}\\
    % %  \sum_{j\in \mathcal{I}_{i,\lambda}}q_{i,j,\lambda}\rho_j\left(k\right)
    %  \rho_i\left(k\right)&i\in \mathcal{V}_{out}\\
    %  \sum_{l=0}^{n_\tau-1}p_{i,\lambda[k-l]}\rho_{i}[k-l]&i\in \mathcal{V}\setminus  \mathcal{V}_{out}
    %  \end{cases}
 \end{equation}
%  \jb{does that man $\rho_i(k+1)=0$ for $i\in\mathcal{V}_E$?}\hrs{Yes}\jb{but then for the previous time we also get $\rho_i(k)=0$, don't we? Hence $y_i(k)=z_i(k)$ at every step, is that correct?}
 \begin{equation}
     y_{i}\left(k\right)=
     \begin{cases}
     u_i(k)&i\in \mathcal{V}_{in},~ \forall \lambda\in \mathbf{\Lambda}\\
     \sum_{j\in \mathcal{I}_{i,\lambda}}q_{i,j}\left(\lambda\right)z_{j}\left(k\right)+d_i&i\in \mathcal{V}\setminus \mathcal{V}_{in},~ \forall \lambda\in \mathbf{\Lambda}
    %  \begin{cases}
    %  u_{e_i}\left(k\right)&e_i\in \mathcal{V}_{in}\\
    %  \sum_{e_j\in \mathcal{I}_{i,\lambda}}q_{i,j,\lambda}Z_{j}\left(k\right)&e_i\in \mathcal{V}\setminus \mathcal{V}_{in}\\
 \end{cases}
    %  ,\qquad 
 \end{equation}
 \end{subequations}
%  \jb{Equation (12b) doesn't make sense, $y_i$ appears on both sides.}
and inflow $y_i\geq 0$ at road element $i\in \mathcal{V}_{in}$  has the following properties:
\begin{enumerate}
     \item{If $i\in \mathcal{V}_{in}$, $y_i=u_i$ can be controlled by a ramp meter.}
     \item{If $i\in \mathcal{V}_{I}$, $d_i\geq 0$ is given as a non-zero-mean Gaussian process.}
\end{enumerate}
  Note that $d_i$ is uncontrolled at road element $i\in \mathcal{V}_R\setminus \mathcal{V}_{in}$. Variable $p_{i}(\lambda)\in \left[0,1\right]$ is the traffic outflow probability, and $q_{i,j}\left(\lambda\right)$ is the outflow fraction of road element $j$ directed towards $i\in \mathcal{O}_{j,\lambda}$ when $\lambda\in \mathbf{\Lambda}$ is the active movement phase over time interval $[t_k,t_{k+1}]$. Note that 
  \begin{equation}
      \sum_{j\in \mathcal{O}_{i,\lambda}}q_{j,i}\left(\lambda\right)=1
  \end{equation}
 for every $\lambda\in \mathbf{\Lambda}$. 
We define $\mathbf{P}\left(\lambda\right)=\mathrm{diag}\left(p_{1}\left(\lambda\right),\cdots,p_{N}\left(\lambda\right),p_{N+1}\left(\lambda\right)\right)$, where $p_{N+1}\left(\lambda\right)=0$  $\forall \lambda\in\Lambda$. This implies that the outflow of the exit node is zero. Also, matrix $\mathbf{Q}(\lambda)=\left[q_{i,j}\left(\lambda\right)\right]\in \mathbb{R}^{\left(N+1\right)\times \left(N+1\right)}$ is non-negative, and 
 \begin{equation}
 \label{qno1}
     q_{N+1,j}\left(\lambda\right)=
     \begin{cases}
     1&j=N+1\in \mathcal{V}_E\\
     0&\mathrm{otherwise}
     \end{cases}
     .
 \end{equation}
Eq. \eqref{qno1} implies that traffic does not flow from the exit node $N+1\in \mathcal{V}_E$ to any other element $j\in \mathcal{V}_R\setminus \mathcal{V}_E$.
The traffic network dynamics is given by
 \begin{equation}
 \label{MPCTrafficDynamics}
    \mathbf{x}\left(k+1\right)
     =\mathbf{A}\left(\lambda\right)
    \mathbf{x}\left(k\right)+\mathbf{g}\left(k\right)
     \end{equation}
where $\mathbf{x}\left(k\right)=\begin{bmatrix}\rho_{1}\left(k\right)&\cdots&\rho_{N+1}\left(k\right)\end{bmatrix}^T$
and $\mathbf{g}=\begin{bmatrix}
{\mathbf{g}}_R^T&g_{N+1}
\end{bmatrix}^T=
\left[g_i\right]\in \mathbb{R}^{\left(N+1\right)\times 1}$ is defined as follows:
\begin{equation}
\label{yiiii}
    g_i(k)=
    \begin{cases}
    u_i(k)&i\in\mathcal{V}_{in}\\
    d_i(k)&i\in\mathcal{V}_R\setminus \mathcal{V}_{in}\\
    0&i\in \mathcal{V}_E
    \end{cases}
    .
\end{equation}
Also, 
\[
 \mathbf{A}\left(\lambda\right)
     =\mathbf{I}-\mathbf{P}\left(\lambda\right)+\mathbf{Q}\left(\lambda\right)\mathbf{P}\left(\lambda\right)=
     \begin{bmatrix}
     \mathbf{\mathbf{C}}\left(\lambda\right)&\mathbf{0}\\
     \mathbf{D}\left(\lambda\right)&1
     \end{bmatrix}
 ,
\]
% \jb{what is new? Isn't this the same as DSCC?}\hrs{No, this is not the same as DSCC paper. Here sum of every column of matrix $\mathbf{A}$ is $1$ and matrix $\mathbf{A}$ is Markov. Furthermore, $\mathbf{A}$ is a function of $\lambda\in \mathbf{\Lambda}$ and $\lambda$ is a finite set.}
where every column of non-negative matrix $\mathbf{A}:\mathbf{\Lambda}\rightarrow \mathbb{R}^{\left(N+1\right)\times \left(N+1\right)}$ sums to $1$ for every movement phase $\lambda\in \Lambda$, $\mathbf{C}:\mathbf{\Lambda}\rightarrow  \mathbb{R}^{N\times N}$, and $\mathbf{D}\left(\lambda\right)\in \mathbb{R}^{1\times N}$. Eigenvalues of matrix $\mathbf{C}\left(\lambda\right)$ are all placed inside a disk of radius $r_{\lambda}<0$ with center at the origin. Note that the $i$-th entry of matrix $\mathbf{D}:\mathbf{\Lambda}\rightarrow \mathbb{R}^{1\times N}$ specifies the fraction of traffic flow exiting the NOIR from node $i\in \mathcal{V}_R$. Traffic dynamics at non-exit nodes is given by
\begin{equation}
\label{rawtrafficdynamics}
    \mathbf{x}_R\left(k+1\right)=\mathbf{C}\left(\lambda\right)\mathbf{x}_R\left(k\right)+\mathbf{g}_R\left(k\right),
\end{equation}
where $\mathbf{x}_R\left(k\right)=\begin{bmatrix}
\rho_1\left(k\right)&\cdots&\rho_N\left(k\right)
\end{bmatrix}^T$.

\section{\hspace{0.2cm} Problem Specification}\label{Problem Formulation}
Linear Temporal Logic (LTL) is used to specify the conservation-based traffic coordination dynamics  \cite{wongpiromsarn2009receding} and present the feasibility conditions. Every LTL formula consists of a set of atomic propositions, logical operators, and temporal operators. Logical operators include $\lnot$ (``negation''), $\vee$ (``disjunction''), $\wedge$ (``conjunction''), and $\Rightarrow$ (``implication''). LTL formulae also use temporal operators  $\square$ (``always''), $\bigcirc$ (``next''), $\lozenge$ (``eventually''), and $\mathcal{U}$ (``until''). 

We extend discrete-time LTL with the syntactic sugar $\Box_{\{0,...,N_\tau\}}$ to specify satisfaction of a certain property in the next $N_\tau+1$ time steps. More specifically,  $\Box_{\{0,...,N_\tau\}} \varphi$  at discrete time $k$ if and only if $\varphi$ is satisfied at discrete times  $k$ to time $k+N_\tau$ \cite{DSCC2020}. 

% \jb{What in this part is new compared to the DSCC paper? These first sentences seem copy-pasted from the DSCC paper, including the new notation. Please make it vey clear this was already done in the DSCC paper by citing it, and point out the exact differences.}\hrs{The main novelty is that traffic coordination is modeled by a switching dynamics with continuous state $\mathbf{x}$ and discrete state $\lambda$. Please read the entire Section. The continuation of this Section is totally new.}
% In discrete-time LTL, the operator $\Box_{\{0,...,N_\tau\}}$ can be defined syntactically as:
% \nv
% $\Box_{\{0,...,N_\tau\}}\varphi\triangleq
% \varphi\land\bigcirc\varphi\land\bigcirc\bigcirc\varphi\land\ldots\land\underbrace{\bigcirc\cdots\bigcirc}_{N_\tau\text{ times}}\varphi\nv$
% The notation is inspired by Metric Temporal Logic MTL~\cite{koymans1990specifying} and Signal Temporal Logic STL~\cite{maler2004monitoring}, which feature a similar operator in continuous time. 

The problem of traffic coordination can be formally specified by a finite-state abstraction defined by tuple
% paper defines a traffic coordination problem by tuple
\[
\mathrm{M}=\left(\mathcal{S},\mathcal{A},\mathcal{H},\mathrm{C}\right),
\]
where $\mathcal{S}$ is the state set, $\mathcal{A}$ is the discrete action set, $\mathcal{H}:\mathcal{S}\times \mathcal{A}\rightarrow \mathcal{S}$
is the state transition relation, and $\mathrm{C}:\mathcal{S}\times \mathcal{A}\rightarrow \mathbb{R}_+$ is the immediate cost function.

%\subsection{State Set $\mathcal{S}$}
% where
% \begin{equation}
%     f_i=f_{min}+f_{\mathrm{min}}+(i-1)\Delta f, 
% \end{equation}
%  $i=\left[{f-f_{\mathrm{min}}\over \Delta f}\right]\in \{1,\cdots,n_f\}$ assigns a discrete value for ${f}\left(\mathbf{g},\lambda,\theta\right)$.

% Define the activation index set 
% \begin{equation}
%     \mathcal{T}=\{\tau=\left(\tau_{1},\cdots,\tau_{m}\right)\in \{0,1\}^m\big|\tau_i\in \{0,1\},~i\in \mathcal{W}\}
% \end{equation}
% to assign all possible activation indexes across the NOIR junctions. 

\subsection{State set $\mathcal{S}$}
Set $\mathcal{S}$ is mathematically defined by
\begin{equation}
    \mathcal{S}=\{s=\left(\mathbf{x},\mathbf{g},\lambda,\tau\right)\big|\mathbf{x}\in\mathbf{X},~\mathbf{g}\in\mathbf{G},~\lambda\in \mathbf{\Lambda},~\tau\in \{0,1\}^m\},
\end{equation}
% where 
% $\mathbf{E}\subset \mathbb{R}^{N+1}$ is a compact set. 
% The state set is then defined by
% \[
% \mathcal{S}=\mathbf{E}\times{\Lambda}\times \mathcal{T}.
% % =\{\left(i,\lambda_{_{1,\zeta_{_{1}}}},\cdots,\lambda_{_{m,\zeta_{_{m}}}},\tau_{1},\cdots,\tau_m\right)\big|\}
% \]
% where 
% ${s}=\left(\mathbf{e},\lambda,\tau\right)\in \mathcal{S}$, 
where the traffic density vector $\mathbf{x}=\begin{bmatrix}
    \rho_1&\cdots&\rho_N
\end{bmatrix}^T\in \mathbb{R}^{N+1}$ and input vector $\mathbf{g}\in \mathbf{G}\in \mathbb{R}^{N_{in}\times 1}$ were introduced in Section  \ref{Problem Statement}, and $\mathbf{X}$ and $\mathbf{G}$ are compact sets. Also, $\lambda=\left(\lambda_{_{1,\zeta_{_{1}}}},\cdots,\lambda_{_{m,\zeta_{_{m}}}}\right)\in \mathbf{\Lambda}$ is a movement phase, and $\tau= \left(\tau_{1},\cdots,\tau_m\right)\in \{0,1\}^m$ where $\tau_i\in \{0,1\}$ is the activation index at junction $i\in \mathcal{W}$. An execution of the proposed system is expressed by $s=s_0s_1s_2,\cdots$ where $s_k=\left(\mathbf{x}[k],\mathbf{g}_k,\lambda[k],\tau[k]\right)$ is the state of the system at time $k$.

\textbf{Feasibility Condition 1:} Traffic density, defined as the number of cars at a road element, is a positive quantity everywhere in the NOIR. It is also assumed that every road element has  maximum capacity $\rho_{\mathrm{max}}$. Therefore, the number of cars cannot exceed $\rho_{\mathrm{max}}$ in any road element $i\in \mathcal{V}$.
% \jb{Do you mean $i\in\mathcal{V}$? Otherwise the formula below is inconsistent}
These two requirements can be formally specified as follows:
\begin{equation}
\label{deltaineq}
    % \forall i\in \mathcal{V},\qquad 
   \displaystyle\bigwedge_{i\in \mathcal{V}}\displaystyle\Box_{\{0,...,N_\tau\}}\left(\rho_i\geq0\ \wedge\  \rho_i\leq\rho_{\mathrm{max}}\right).\tag{$\Phi_1$}
    % \fbox{$\forall k,~\forall i\in \mathcal{V},\qquad 0\leq \rho_i[k]\leq \rho_{\mathrm{max}}$}
\end{equation}
%\jb{Why only up to time $N_\tau$? Why not forever?}\hrs{because the optimization problem is for the next $N_\tau$ steps.}
If feasibility condition $\Phi_1$ is satisfied at every road element, then traffic over-saturation is avoided everywhere in the NOIR, at every discrete time $k$.

% \noindent
\textbf{Optional Condition 2:} Boundary inflow should satisfy the following feasibility condition at every discrete time $k$:
% \nv
\begin{equation}
\label{deltaineqequality}
    % \forall i\in \mathcal{V},\qquad 
    \Box_{\{0,...,N_\tau\}}\left(\displaystyle\sum_{i\in \mathcal{V}_{in}}{u}_i=u_0\right).\tag{$\Phi_{2}$}
\end{equation}
Boundary condition \eqref{deltaineqequality} constrains the number of vehicles entering
the NOIR to be exactly $u_0$ at any time $k$. { Note that $u_0$ is an upper bound on vehicles entering the NOIR.
%\jb{Shouldn't you write $\displaystyle\sum_{i\in \mathcal{V}_{in}}{u}_i\leq u_0$ then?}\hrs{The simulation is based on $\Phi_2$. Thus $\Phi_2$ is correct.}
However, in the simulation results presented, traffic demand is significant such that the NOIR is maximally utilized by as many vehicles as possible.}

% \textbf{Atomic propositions} are described using linear temporal logic (LTL). In particular, this paper uses LTL to specify the feasibility conditions of the conservation-based traffic coordination dynamics given in equation~\eqref{MAIN} \cite{wongpiromsarn2009receding}. Every LTL formula consists of a set of atomic propositions, logical operators, and temporal operators. Logical operators include $\lnot$ (``negation''), $\vee$ (``disjunction''), $\wedge$ (``conjunction''), and $\Rightarrow$ (``implication''). LTL formulae also use temporal operators  $\square$ (``always''), $\bigcirc$ (``next''), $\lozenge$ (``eventually''), and $\mathcal{U}$ (``until'').  set $\mathcal{P}$ is defined as $\mathcal{P}=\left\{\right\}$, where atomic propositions $\psi_1$ through $\psi_8$ are defined as follows:

% \begin{subequations}
% \begin{equation}
%     \bigwedge_{i\in \mathcal{V}}\left(\rho_i\geq  \rho_{\mathrm{max}}\right), \tag{$\psi_{1,i}$}
% \end{equation}
% \begin{equation}
%     \bigwedge_{i\in \mathcal{V}}\left(\rho_i\leq \rho_{\mathrm{max}}\right), \tag{$\psi_{2,i}$}
% \end{equation}
% \begin{equation}
%   \bigwedge_{j\in \mathcal{W}} \left(\left(\lambda_j,\bigcirc \lambda_j\right)\in \mathcal{C}_j\right), \tag{$\psi_1$}
% \end{equation}
% \begin{equation}
%   \bigwedge_{j\in \mathcal{W}} \left(\bigcirc \lambda_j=\lambda_j\right), \tag{$\psi_2$}
% \end{equation}
% \begin{equation}
%   j\in \mathcal{W},\qquad  \tau_j=1, \tag{$\psi_{3,j}$}
% \end{equation}
% \end{subequations}

\subsection{Action Set $\mathcal{A}$}
Action set $\mathcal{A}: \mathbf{\Lambda}\times \mathcal{T}\rightarrow \mathbf{\Lambda}$ assigns the next acceptable movement at every junction $i\in \mathcal{W}$, given the current NOIR activation index $\tau\in \mathcal{T}=\{0,1\}^m$ and movement phase $\lambda=\left(\lambda_1,\cdots,\lambda_m\right)$, {i.e. $\tau=\left(\tau_1,\cdots,\tau_m\right)$, $\tau_i\in \{0,1\}$, $i\in \mathcal{W}$}.
{We write $\lambda_i^+$ for the value of $\lambda_i$ in the next state, i.e. $\lambda_i^+(k)=\lambda_i(k+1)$, and similarly for other variables.
}
Actions are constrained and must satisfy one of the following LTL formula: 
\begin{subequations}
\label{tau}
\begin{equation}
\label{tau1}
  \left(\tau_i=0\right)\Rightarrow\left(\left(\lambda_i, \lambda_i^+\right)\in \Xi_i\vee \left(\lambda_i^+=\lambda_i\right)\right),\tag{$\Phi_{3,i}$}
\end{equation}
\begin{equation}
\label{tau2}
\left(\tau_i=1\right)\Rightarrow\left(\lambda_i, \lambda_i^+\right)\in \Xi_i,
%   \left(\tau_i=1\right)\Rightarrow\left(\bigcirc \lambda_i=\lambda_i\right),
\tag{$\Phi_{4,i}$}
\end{equation}
\end{subequations}
Combining \eqref{tau1} and \eqref{tau2}, the next movement phase must satisfy the following LTL formula:
\begin{equation}
\label{tau4}
  \bigwedge_{i\in \mathcal{W}}\Box_{\{0,...,N_\tau\}}\left(\left(\left(\left(\lambda_i,\lambda_i^+\right)\in \Xi_i\right)\mathcal{U}\left(\tau_i=1\right) \right)\vee \left(\left(\lambda_i,\lambda_i^+\right)\in \Xi_i\right)\right).\tag{$\Phi_5$}
\end{equation}
\jb{Do you see this equation as a consequence/equivalent formulation to the previous two? If so I don't understand where the Until comes from.}
\begin{remark}
Set $\mathcal{A}(\lambda,\tau)\subset \mathbf{\Lambda}$ is defined as follows:
% \begin{equation}
%     \mathcal{A}\left(\lambda,\tau\right)=\left\{a\left(\lambda,\tau\right)=\lambda^+: a\left(\lambda,\tau\right)\models \Phi_5 \right\},
% \end{equation}
\begin{equation}
    \mathcal{A}(\lambda,\tau)=\{\lambda^+\in\mathbf{\Lambda}\ |\ \forall i\in \mathcal{W}, (\lambda_i,\lambda_i^+)\in\Xi_i\lor(\tau_i=0\land\lambda_i^+=\lambda_i)\}.
\end{equation}
%where $a\left(\lambda,\tau\right)=\left(\lambda_1^+,\cdots,\lambda_m^+\right)$.
\end{remark}

\subsection{State Transition Function} 
 The state transition relation $\mathcal{H}$ defines transition from ``current'' state $s=\left(\mathbf{x},\mathbf{g},\lambda,\tau\right)\in \mathcal{S}$ to ``next'' state  $ s^+=\left(\mathbf{x}^+,\mathbf{g}^+,\lambda^+,\tau^+\right)\in \mathcal{S}$ given action $a\left(\lambda,\tau\right)\in \mathcal{A}\left(\lambda,\tau\right)$. \jb{where/why do you need $a$? It shows up nowhere in the equations beforehand}
% Transition over the state space space is partially determninistic and partially probabilistic. 
Current and next movement phases must satisfy condition \eqref{tau5} below. 

Transition of current activation index $\tau$ must satisfy the following properties:
\begin{equation}
\label{tau5}
  \bigwedge_{i\in \mathcal{W}}\left(\left( \tau_i^+=0\right)\mathcal{U}\left(T_i=T_{L,i}\right)\right).\tag{$\Phi_6$}
\end{equation}
\jb{Isn't this true by definition of $\tau_i$ and the fact that time always moves forward? Why do you need it?} Note that the $T_i$ is reset every time movement phase is updated at junction $i\in \mathcal{W}$. This requirement is formally specified as follows:
\begin{equation}
\forall i\in \mathcal{W},\qquad     \left( \lambda_i^+\neq \lambda_i\right)\Rightarrow \left(T_i^+=0\right)
\end{equation}
This paper assumes that ${g}_i=d_i$ is a Gaussian process for $i\in \mathcal{V}_I$ is an non-inlet road, i.e. $d_i\sim\mathcal{N}\left(\bar{{d}}_i,{\sigma}_i\right) $. Per Eq. \eqref{yiiii}, $g_i=u_i$ for $i\in \mathcal{V}_{in}$ where $u_i$ is determined as the solution of a receding horizon optimization problem presented in Section \ref{Traffic Coordination Control}. Therefore
\begin{equation}
% \left(g_i^+= u_i^+\right)\vee\left( g_i^+=y_i\right)\vee\left(g_i^+=0\right).
    \left(\bigwedge_{i\in \mathcal{V}_{in}}g_i^+= u_i^+\right)\wedge\left(\bigwedge_{i\in \mathcal{V}_I} g_i^+=y_i\right)\wedge\left(\bigwedge_{i\in \mathcal{V}_E}g_i^+=0\right).
\end{equation}
% \jb{Why $\lor$ in between the three terms? Did you mean $\land$?}

Transition of $\mathbf{x}$ is governed by \eqref{MPCTrafficDynamics}, thus
\begin{equation}
    \mathbf{x}^+=\mathbf{A}\left(\lambda\right)\mathbf{x}+\mathbf{g}
\end{equation}
where $\lambda\in \mathbf{\Lambda}$.

\subsection{Cost Function}
\jb{Why is this Subsection in Section IV? It has nothing to do with specification}
%  \begin{remark}
 Given Eq. \eqref{MPCTrafficDynamics}, an $N_\tau$-step expected transition is given by
 \begin{equation}
 \mathbf{x}_{N_\tau+1}=\mathbf{\Theta}_h\left(\lambda\right)\mathbf{x}_1+\mathbf{\Gamma}_{N_\tau}
 \begin{bmatrix}
 \mathbf{g}_1\\
 \vdots\\
 \mathbf{g}_{N_\tau}
 \end{bmatrix}
 ,
 \end{equation}
 where  $\mathbf{g}_1,\cdots,\mathbf{g}_{N_\tau}\in \mathbf{G}$, $\mathbf{x}_1\in \mathbf{X}$,  $\lambda\in \mathbf{\Lambda}$,
 \[
 \mathbf{\Theta}_{N_\tau}\left(\lambda\right)=\mathbf{A}^{N_\tau}\left(\lambda\right)
%  _{N_\tau}\right)\cdots\mathbf{A}\left(\lambda_{N_\tau}\right)\in \mathbb{R}^{\left(N+1\right)\times \left(N+1\right)},
 \]
 and
 \[
 \mathbf{\Gamma}_{N_\tau}=
 \begin{bmatrix}
 \mathbf{\Theta}_{N_\tau-1}&\cdots&\mathbf{\Theta}_1&\mathbf{I}
 \end{bmatrix}
 \in \mathbb{R}^{\left(N+1\right)\times N_\tau\left(N+1\right)}.
 \]
%  \end{remark}
 
The cost function $\mathrm{C}$ is defined by
 \begin{equation}
 \label{Costfucction}
\begin{split}
     \mathrm{C}\left(\mathbf{x},\mathbf{g}_1,\cdots,\mathbf{g}_{N_\tau},\lambda\right)=&\sum_{h=1}^{N_\tau}\mathbf{x}_{h+1}^T\mathbf{F}^T\mathbf{F}\mathbf{x}_{h+1}\\
     =&\begin{bmatrix}
     \mathbf{x}_1^T&\mathbf{g}_1^T&\cdots&
      \mathbf{g}_{N_\tau}^T
     \end{bmatrix}
    \mathbf{W}
     \begin{bmatrix}
     \mathbf{x}_1\\
     \mathbf{g}_1\\
     \vdots\\
      \mathbf{g}_{N_\tau}\\
     \end{bmatrix}
\end{split}
\end{equation}
where   
\[
\mathbf{F}=\begin{bmatrix}\mathbf{I}_{N}&\mathbf{0}_{N\times 1}
\\\mathbf{0}_{1\times N}&0
\end{bmatrix},
\]
and
\[
\mathbf{W}= \begin{bmatrix}
     \sum_{h=1}^{N_\tau}\mathbf{\Theta}_h^T\mathbf{F}^T\mathbf{F}\mathbf{\Theta}_h&\sum_{h=1}^{N_\tau-1}\mathbf{\Theta}_{N_\tau}^T\mathbf{F}^T\mathbf{F}\mathbf{\Gamma}_h\\
     \sum_{h=1}^{N_\tau}\mathbf{\Gamma}_h^T\mathbf{F}^T\mathbf{F}\mathbf{\Theta}_{N_\tau}&\sum_{h=1}^{N_\tau}\mathbf{\Gamma}_h^T\mathbf{F}^T\mathbf{F}\mathbf{\Gamma}_{h}
     \end{bmatrix}
     .
\]
% \subsection{Traffic Feasibility Conditions}
% We use linar temporal logic to specify the traffic feasiblity conditions. 

\section{Traffic Congestion Control}
\label{Traffic Coordination Control}
The objective of the traffic congestion control is to determine optimal inflow
and movement phase such that cost function $\mathrm{C}$, defined in Eq.  \eqref{Costfucction}, is minimized. Optimal traffic inflow is assigned with MPC while optimal movement phases are assigned as the solution of a RHO problem.

The optimal boundary inflow $\mathbf{g}_1^*$ is assigned by solving the following optimization problem:
\begin{equation}
\mathbf{x}\in \mathbf{X},~\lambda\in\mathbf{\Lambda},~~ \left(\mathbf{g}_1^*,\cdots,\mathbf{g}_{N_\tau}^*\right)=\argmin\limits_{\mathbf{g}_1,\cdots,\mathbf{g}_{N_\tau}\in \mathbf{G}}~\mathrm{C}\left(\mathbf{x},\mathbf{g}_1,\cdots,\mathbf{g}_{N_\tau},\lambda\right),
\end{equation}
subject to the conditions \eqref{deltaineq} and \eqref{deltaineqequality}.
The optimal movement phase $\lambda^{+*}$ is assigned by solving the following optimization problem:
\begin{equation}
\begin{split}
    &\mathbf{x}\in \mathbf{X},~\mathbf{g}_1,\cdots,\mathbf{g}_{N_\tau}\in \mathbf{G},~\lambda\in \mathbf{\Lambda},\\ &\lambda^{+*}=\argmin\limits_{\lambda^+\in \mathcal{A}\left(\lambda,\tau\right)}\mathrm{C}\left(\mathbf{x},\mathbf{g}_1,\cdots,\mathbf{g}_{N_\tau},\lambda\right),
\end{split}
\end{equation}
subject to the following conditions $\bigwedge_{i\in \mathcal{W}}\Phi_{i,3}$, $\bigwedge_{i\in \mathcal{W}}\Phi_{i,4}$, and $\Phi_{5}$.
% \[
% \lambda^{h+1}\in \mathcal{A}\left(\lambda^h,\tau(h)\right),
% \]
% for $h=1,\cdots,N_\tau-1$, where $\tau(h)\in \mathcal{T}$.
\section{Simulation Results}\label{Simulation Results}
Traffic coordination is investigated in simulation for the example NOIR shown in Fig. \ref{NOIRExample} (a)  consisting of $N=53$ unidirectional roads. Traffic coordination is controlled through the NOIR inlet boundary nodes defined by $\mathcal{V}_{in}=\{1,\cdots,8\}$ and traffic signals at junction nodes $\mathcal{W}=\{1,\cdots,17\}$. 
% Note that single roads is incident at junct

This paper assumes that the time interval between two consecutive discrete times $k$ and $k+1$ is $\Delta t=30s$. 
It is assumed that the inflow $y_i={\frac{1}{2}}\pm0.5$ is randomly entered through every road element $i\in \mathcal{V}_I$. For simulation $u_0=31$ is chosen. Therefore, a total of $31$ vehicles are allowed to enter the NOIR through the NOIR inlet boundary road elements at every discrete time $k$. Traffic coordination is controlled through the ramp meter at the NOIR boundary road elements and traffic signals at NOIR intersections by solving the optimization problem developed in Section \ref{Traffic Coordination Control}. 

\begin{figure}
\center%{\epsfig{figure=fig1.eps,width=6.85in}}
\includegraphics[width=3.2 in]{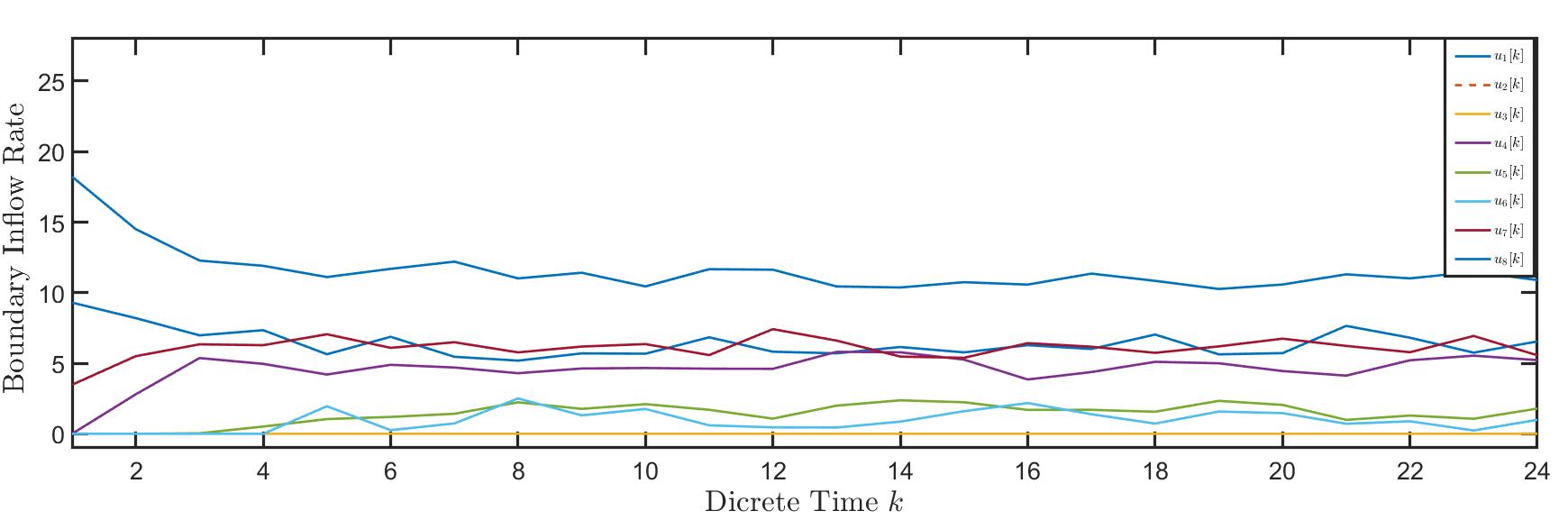}
\caption{Optimal boundary inflow rates $u_1$ through $u_8$ versus discrete time $k$.}
\label{BoundaryInput}
\end{figure}

\begin{figure}
\center%{\epsfig{figure=fig1.eps,width=6.85in}}
\includegraphics[width=3.2 in]{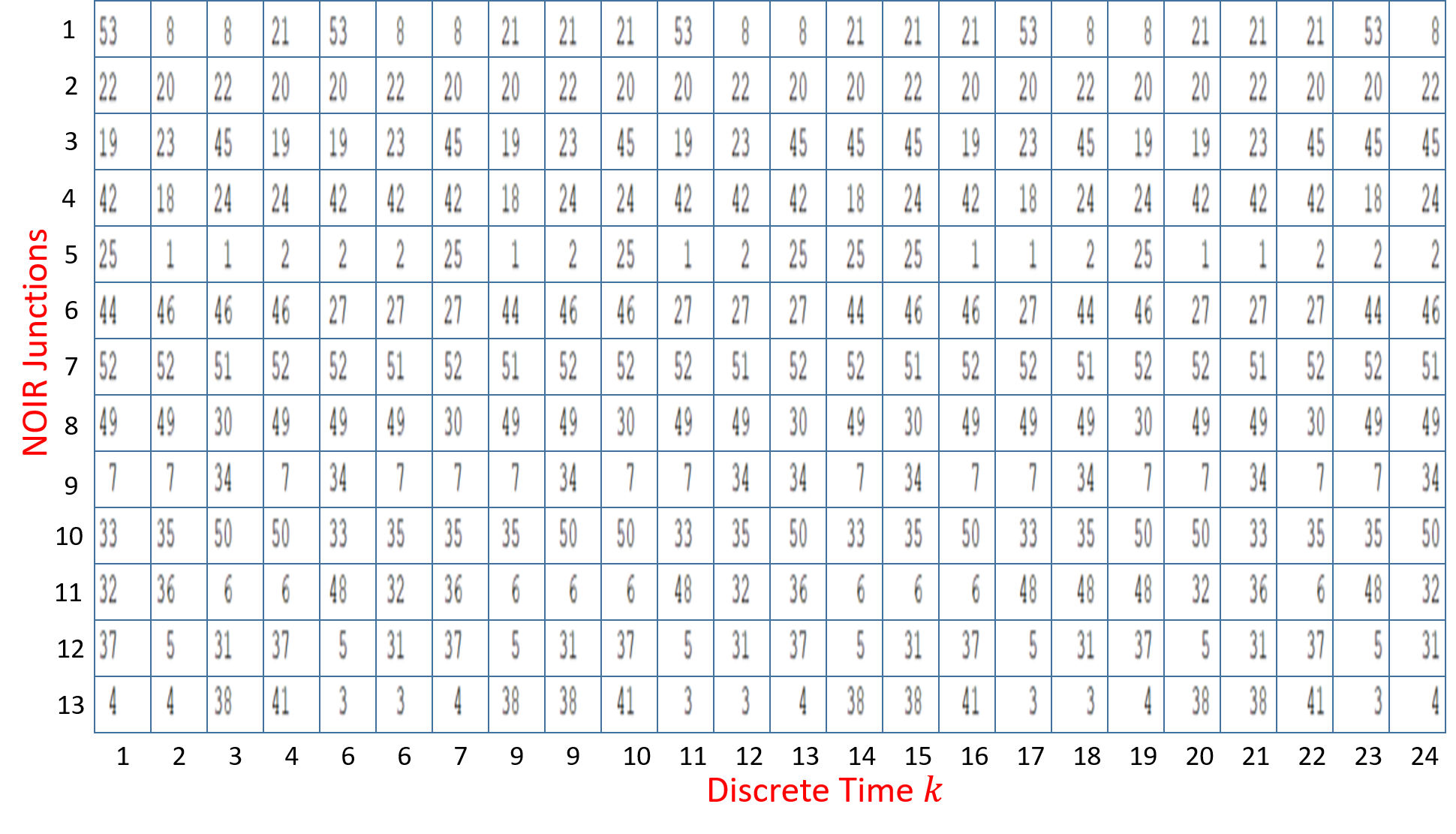}
\caption{Optimal movement phases at NOIR junctions at $k=1,\cdots,24$.}
\label{OptimalActions}
\end{figure}

\begin{figure}
\center%{\epsfig{figure=fig1.eps,width=6.85in}}
\includegraphics[width=3.2 in]{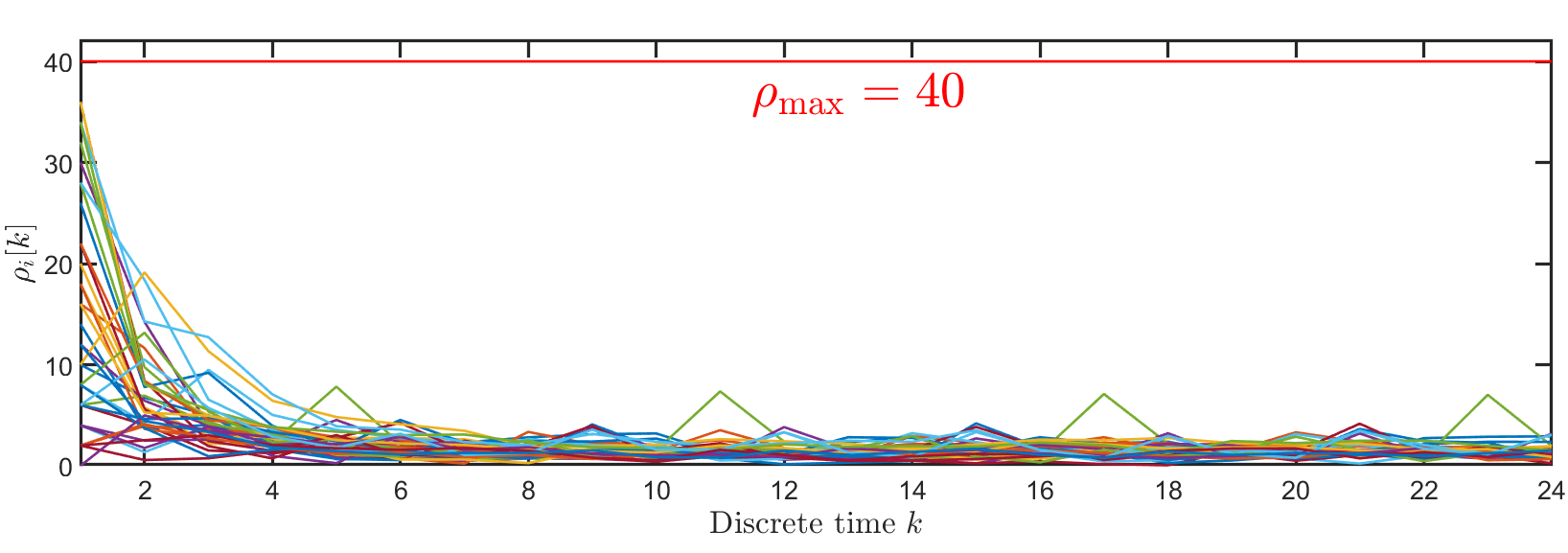}
\caption{Traffic density at every NOIR road for $k=1,\cdots,24$.}
\label{trafficdensityalla}
\end{figure}

\begin{figure}
\center%{\epsfig{figure=fig1.eps,width=6.85in}}
\includegraphics[width=3.5 in]{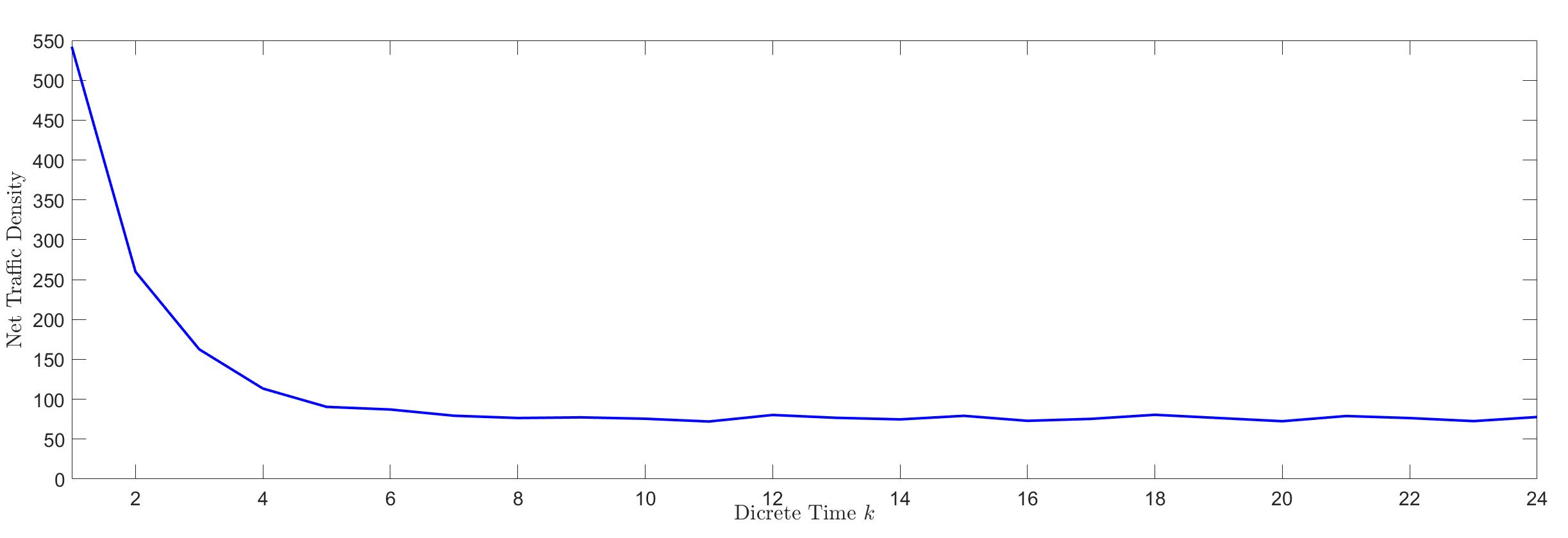}
\caption{Total traffic density of the NOIR vs. discrete time $k$.}
\label{NetTrafficDensity}
\end{figure}

In Fig. \ref{BoundaryInput}, boundary inflow rates $u_1$ through $u_8$ are plotted versus time for $k=1,\cdots,100$. For the simulation, $\rho_{\mathrm{max}}=40$ is considered. Fig.  \ref{trafficdensityalla} plots traffic density $\rho_i$ at every road element $i\in \mathcal{V}$ versus discrete time $k$. It is seen that $\rho_(k)<\rho_{\mathrm{max}}=40$ at every discrete time $k$. Thus, traffic oversaturation is ensured.
Also, the total traffic density $r_{\mathrm{net}}(k)=\mathbf{1}_{1\times N}\mathbf{x}_R(k)$ is plotted versus discrete time $k$ in Fig. \ref{NetTrafficDensity}. For simulation, we choose $T_{L,i}=3$. Therefore, a movement phase cannot be active more than $3\times\Delta T=90s$. A movement phase at junction $i\in \mathcal{W}$ is represented by a directed tree containing a root node and terminal nodes per the example movement phase shown in Fig. \ref{NOIRExample} (b). The root node represents the active road with incoming traffic flow, and terminal nodes represent the active outgoing roads.  In Fig. \ref{OptimalActions}, active incoming roads are shown at NOIR junctions $1,\cdots,13\in \mathcal{W}$ for $k=1,\cdots,24$. 

Fig. \ref{NetTrafficDensity} plots the net traffic density of the NOIR versus discrete time $k$ for $k=1,\cdots,24$. It is seen that net traffic density reaches the steady-state value in about eight time steps while traffic consistently enters and leaves the NOIR.

\section{\hspace{0.2cm}Conclusion}\label{Conclusion}
This paper offers a physics-inspired approach to model and control traffic coordination in a network of interconnected roads (NOIR). Traffic coordination modeled as a Markov process is obtained by spatial and temporal discretization of the mass conservation continuity equation. We showed how traffic congestion can be effectively controlled through ramp meters and traffic signals located at boundaries and junctions of the NOIR. In particular,  MPC is applied to control the boundary inflow while a RHO planner optimizes movement phases commanded by traffic signals at NOIR junctions. Simulation results show that integration of boundary and signal controls can effectively manage urban traffic congestion.
% In particular, we modeled traffic coordination by a Markow process 

\section{\hspace{0.3cm}Acknowledgement}
This work has been supported by the National Science
Foundation under Award Nos. 1914581 and 1739525. The authors gratefully thank Professor Ella Atkins for the useful comments on this paper.

% \textbf{Analysis of the Results:} Let matrix $\mathbf{Q}$

% Therefore, $\mathcal{V}=\mathcal{V}_R\bigcup \mathcal{V}_E$ where $\mathcal{}V_R=\{1,\cdots,53\}$ and $\mathcal{V}_E=\{54\}$ are disjoint sets defining road index numbers and a single exit node, respectively. The NOIR consists 
% Note some features of of state $s=\left(\mathbf{x},\mathbf{u},\lambda,\tau\right)\in \mathcal{S}$ 

\bibliographystyle{IEEEtran}
\bibliography{reference}

\end{document}